\documentclass[fleqn,usenatbib]{mnras}
\pdfoutput=1

\usepackage{newtxtext,newtxmath}

\usepackage[T1]{fontenc}
\usepackage{ae,aecompl}

\usepackage{graphicx}	
\graphicspath{{figures/}}
\usepackage{amsmath}	
\usepackage{multicol}
\usepackage{physics}
\usepackage{siunitx}
\usepackage{tikz}
\usepackage{xspace}

\newcommand{\set}[1]{\left\{#1\right\}}  
\newcommand{\siII}{Si~II~$\lambda$6355\xspace}

\newcommand{\unimodalmu}{\SI{11500}{\km\per\s}}
\newcommand{\unimodalsigma}{\SI{1300}{\km\per\s}}

\newcommand{\bimodalmuI}{\SI{11000}{\km\per\s}}
\newcommand{\bimodalsigmaI}{\SI{700}{\km\per\s}}

\newcommand{\bimodalmuII}{\SI{12300}{\km\per\s}}
\newcommand{\bimodalsigmaII}{\SI{1800}{\km\per\s}}

\newcommand{\caseItheta}{\SI{120(15)}{\degree}}
\newcommand{\caseIdeltav}{\SI{5400(1200)}{\km\per\s}}

\newcommand{\caseIItheta}{\SI{63(6)}{\degree}}
\newcommand{\caseIIdeltav}{\SI{5700(1100)}{\km\per\s}}

\title[Ejecta Velocities of Type Ia Supernovae]{Distribution of \siII Velocities of Type Ia Supernovae and Implications for Asymmetric Explosions}

\author[Zhang et al.]{
Keto D. Zhang,$^{1,2}$\thanks{E-mail: keto.zhang@gmail.com}
WeiKang Zheng,$^{1}$\thanks{E-mail: weikang@berkeley.edu}
Thomas de Jaeger,$^{1,3}$
Benjamin E. Stahl,$^{1,4,5}$
\newauthor{
Thomas G. Brink,$^{1}$
Xuhui Han,$^{6}$
Daniel Kasen,$^{1,4,7}$
Ken J. Shen,$^{1}$
Kevin Tang,$^{1}$
}
\newauthor{
Alexei V. Filippenko$^{1,8}$
}
\\
$^{1}$ Department of Astronomy, University of California, Berkeley, CA 94720-3411, USA. \\
$^{2}$ Google Lick Predoctoral Fellow. \\
$^{3}$ Bengier Postdoctoral Fellow. \\
$^{4}$ Department of Physics, University of California, Berkeley, CA 94720, USA. \\
$^{5}$ Marc J. Staley Graduate Fellow. \\
$^{6}$ Key Laboratory of Space Astronomy and Technology, National Astronomical Observatories, \\
Chinese Academy of Sciences, Beijing 100101, China. \\
$^{7}$ Lawrence Berkeley National Laboratory, Berkeley, CA, USA.\\
$^{8}$ Miller Institute for Basic Research in Science, University of California, Berkeley, CA 94720, USA.
}

\date{Accepted XXX. Received YYY; in original form ZZZ}

\pubyear{2020}

\begin{document}
\label{firstpage}
\pagerange{\pageref{firstpage}--\pageref{lastpage}}
\maketitle

\begin{abstract}
The ejecta velocity is a very important parameter in studying the structure and properties of Type Ia supernovae (SNe~Ia). It is also a candidate key parameter in improving the utility of SNe~Ia for cosmological distance determinations. Here we study the velocity distribution of a  sample of 311 SNe~Ia from the kaepora database. The velocities are derived from the \siII absorption line in optical spectra measured at (or extrapolated to) the time of peak brightness. We statistically show that the observed velocity has a bimodal Gaussian distribution consisting of two groups of SNe~Ia: Group I with a lower but narrower scatter ($\mu_1 = \bimodalmuI$, $\sigma_1 = \bimodalsigmaI$), and Group II with a higher but broader scatter ($\mu_2 = \bimodalmuII$, $\sigma_2 = \bimodalsigmaII$). The population ratio of Group I to Group II is 201:110 (65\%:35\%). There is substantial degeneracy between the two groups, but for SNe~Ia with velocity $v > \SI{12000}{\km\per\s}$, the distribution is dominated by Group II. The true origin of the two components is unknown, though there could be that naturally there exist two intrinsic velocity distributions as observed. However, we try to use asymmetric geometric models through statistical simulations to reproduce the observed distribution assuming all SNe~Ia share the same intrinsic distribution. In the two cases we consider,
35\% of SNe~Ia are considered to be asymmetric in Case 1, and all SNe~Ia are asymmetric in Case 2. Simulations for both cases can reproduce the observed velocity distribution but require a significantly large portion ($>35\%$) of SNe~Ia to be asymmetric. In addition, the Case 1 result is consistent with recent polarization observations that SNe~Ia with higher \siII velocity tend to be more polarized.
\end{abstract}

\begin{keywords}
supernovae: general --- methods: statistical
\end{keywords}



\section{Introduction}

Type~Ia supernovae (SNe~Ia; see, e.g., \citealt{filippenko97} for a review of supernova classification) are the thermonuclear runaway explosions of carbon/oxygen white dwarfs (see, e.g., \citealt{hillebrandt00,howell11} for reviews). One of the most important applications of SNe~Ia is that they can be used as standardisable candles for measuring galaxy distances and thus the expansion history of the Universe (\citealt{riess98,perlmutter99,riess19}).

There are two generally favoured progenitor systems for SNe~Ia:
the single-degenerate scenario (\citealt{whelan73}), in which a single white dwarf accretes material from a nondegenarate companion star, and the double-degenerate scenario
(\citealt{Iben84}; \citealt{webbink84}) involving two white dwarfs.
In the single-degenerate scenario, the nondegenarate companion star
can survive after the SN ejecta collide with the companion star (\citealt{kasen10}).
In the double-degenerate scenario, various models predict different outcomes
for the companion star. For example, in both the merger model \citep[e.g.,][]{pakmor12}
and the head-on collision model \citep[e.g.,][]{kushnir13}, the companion white dwarf is
also destroyed. On the other hand, in some models the companion white dwarf could survive \citep[e.g.,][]{shen18}.
A few SNe~Ia with early-time observations have already ruled out
a giant companion \citep[e.g.,][]{silverman12a, zheng13, goobar14, cao15,
im15, hosseinzadeh17, holmbo19, li19, kawabata19, han20}, and for the extreme case
of SN~2011fe, the companion was constrained to be a white dwarf \citep[][]{nugent11, bloom12},
thus favoring the double-degenerate scenario. However, such early-time observations are still quite rare,
and our understanding of the progenitor systems and explosion mechanisms
remains substantially incomplete both theoretically and observationally
(see a recent review by \citealt{jha19}).

Asymmetric ejecta of SNe~Ia are also predicted by different models. While the single-degenerate
model suggests that the ejecta can be quasispherical on large scales \citep[e.g.,][]{seitenzahl13,sim13}, in violent merger models the ejecta can depart from spherical symmetry
on large angular scales \citep[e.g.,][]{pakmor10,moll14,raskin14},
thus resulting in different ejecta velocities seen from different viewing angles.
The observed ejecta velocity is thus a very important parameter in studying the structure and properties
of SNe~Ia, and can also be used for improving the luminosity vs. light-curve shape relation used to standardise SNe~Ia.
For example, \citet{foley11b}, \citet{wang09,wang13}, and \citet{zheng18} show that by classifying SNe~Ia into subgroups according to their ejecta velocities, or directly adopting ejecta velocity as an additional parameter in the luminosity relation, one can reduce the scatter by 0.04--0.08 mag. 

In this paper, we statistically study the distribution of \siII velocities measured
at the time of peak brightness and use the velocity information as the only input data to
explore the asymmetry  of SNe~Ia through statistical simulations.

\section{Data Selection}
The velocity of the SN ejecta can be measured from the observed absorption minima in P-Cygni line profiles. The most prominent optical line in SNe~Ia during the photospheric phase is \siII, whose velocity is usually an indication of its photospheric velocity. Here we intend to study the \siII velocity distribution at peak brightness; ideally, it is best that a good spectrum be taken right at the time of peak brightness, but this is difficult to do in practice. In general, although the \siII velocity decreases dramatically as a power law during the first few days after explosion, it usually exhibits a slow (close to linear) decline around the time of peak brightness \citep[e.g.,][]{silverman12c,zheng17a,stahl20}. Therefore, as long as there is a spectrum observed within a few days on either side of peak brightness (typically a 1-week interval), one can extrapolate the velocity to the time of peak brightness.

Several groups have already published many \siII velocities near the time of peak brightness \citep[e.g.,][]{foley11a,wang13,folatelli13,zheng18,siebert19}. Among them, \cite{siebert19} released the largest number of appropriate SNe~Ia (311 total) through the kaepora database (v1.0)\footnote{\url{https://msiebert1.github.io/kaepora/}}. This database is compiled from heterogeneous sources (see \citealt[][]{siebert19} and references therein for details), with the majority from the Center for Astrophysics Supernova Program (\citealt{blondin12}), the Berkeley SN~Ia Program (\citealt{silverman12b}), and the Carnegie Supernova Project (\citealt{folatelli13}). It contains all the SNe~Ia originally published by \cite{foley11a}, as well as dozens of new samples measured with the same methods as given by \cite{foley11a}.

\begin{figure}
    \centering
    \includegraphics[width=\columnwidth]{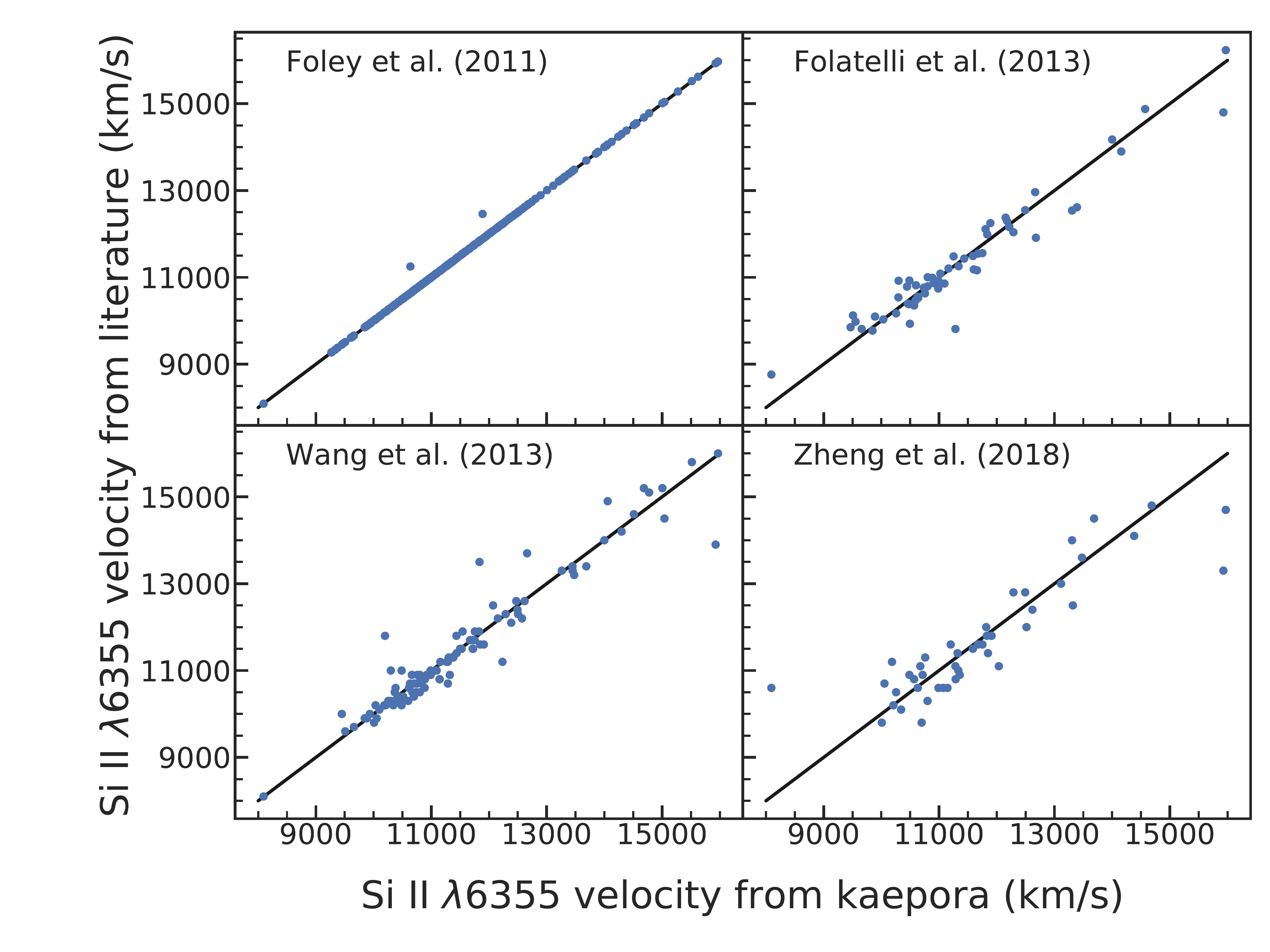}
    \caption{Comparison of \siII velocities of the overlapping SNe between the kaepora sample and the other four sources from the literature.}
    \label{fig:overlap}
\end{figure}

In total, these datasets report the  \siII velocity near the time of peak brightness for 395 SNe~Ia. Note that each research group adopts a different method to measure the \siII velocity, so we compare their values. Since the kaepora database contains the largest sample, we compare other samples with that of kaepora and find an  overlap of 272 objects; see Figure \ref{fig:overlap}. Overall, the velocities are consistent with each other, but a few outliers are seen. To keep the measurements consistent, and also considering that the kaepora sample (311 SNe~Ia) already includes most of the SNe~Ia, we decided to only adopt the 311 objects in the kaepora sample in our analysis.

\begin{figure*}
    \centering
    \includegraphics[width=0.45\linewidth]{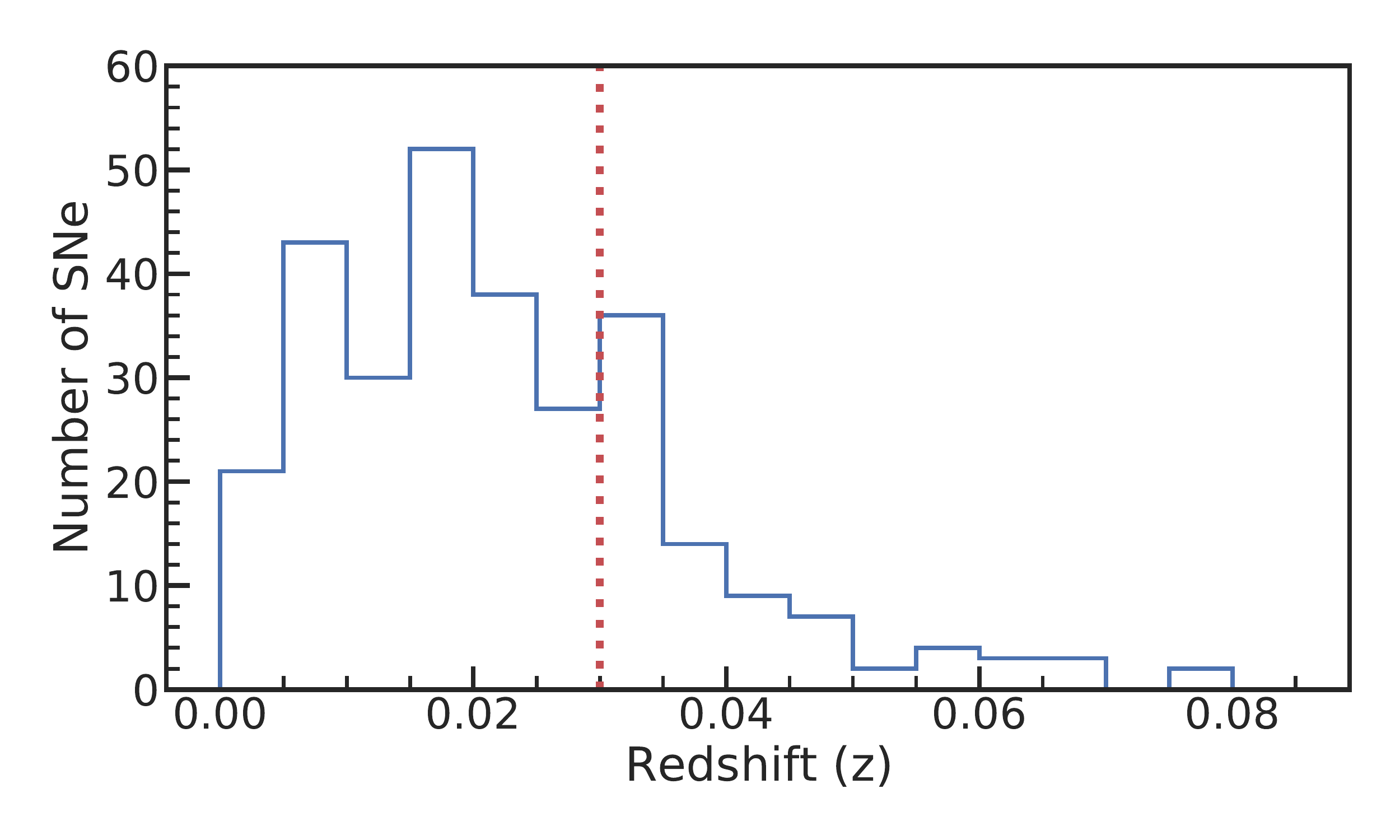}
    \includegraphics[width=0.45\linewidth]{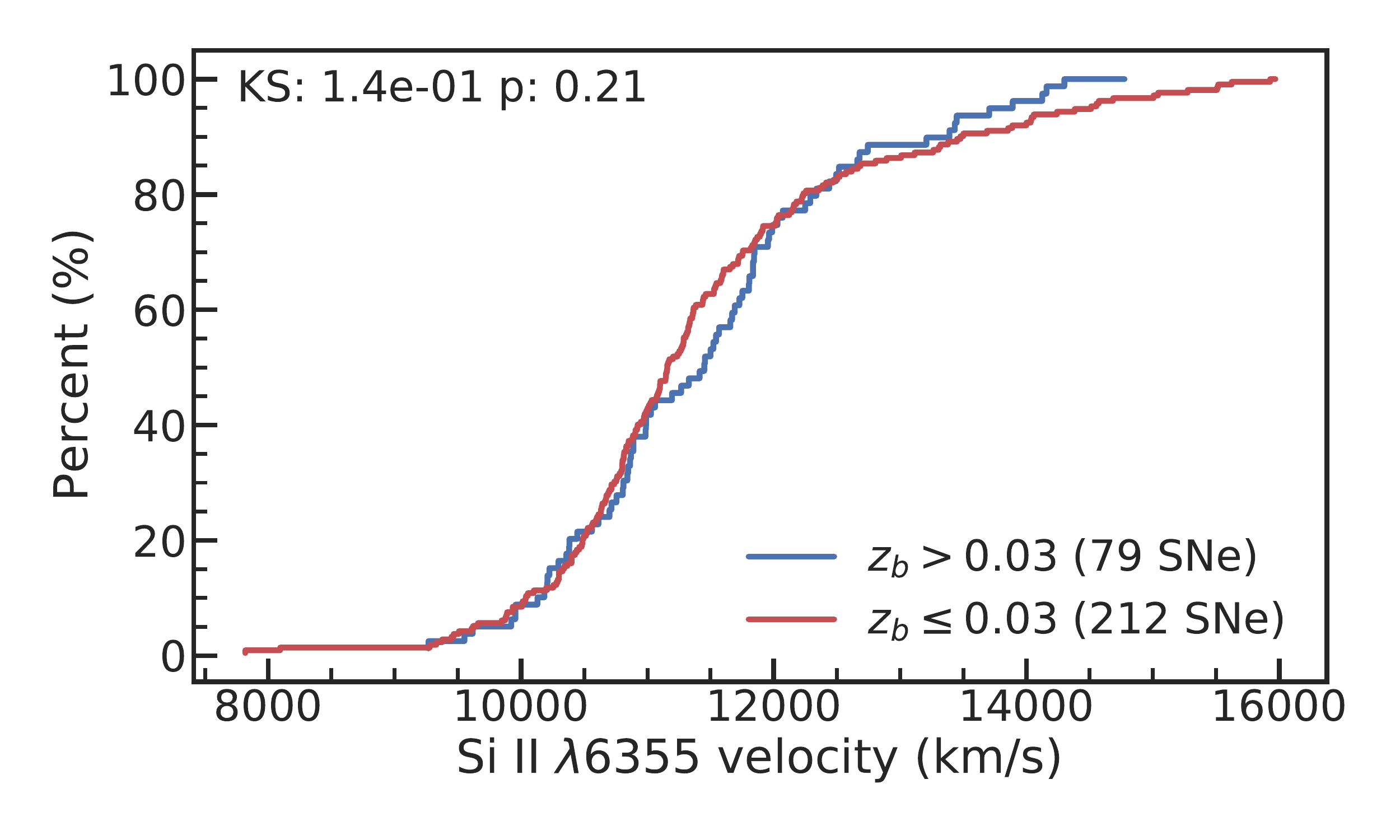}
    \caption{Left: Histogram of 291 SNe~Ia (out of 311) from the kaepora sample that have host-galaxy redshift information, with $z=0.03$ shown as a red dotted line. Right: The cumulative distribution of the 291 SNe~Ia breaks into low- and high- redshift groups with $z_b = 0.03$. A KS test between the two groups gives a $p$-value of 0.21, indicating no significant difference between the two groups.}
    \label{fig:z_histogram}
\end{figure*}

One concern about the data selection is potential bias. Here we select the SNe based on their spectral properties;
specifically, at least one spectrum must be taken within a few days of peak brightness in order to derive the \siII velocity at the time of peak brightness through the relation of \citet[][]{foley11a} (see also \citealt[][]{siebert19}). No other criteria, either the properties of the SN~Ia itself (e.g., ${\bigtriangleup}m_{15}$, rise time) or the properties of its host galaxy (e.g., morphology, mass), are directly used to select the sample; therefore, those properties should not dominate the bias.
However, since the SNe are discovered through photometry, if there is a correlation between the velocity and the peak luminosity, our sample could be biased.
For example, if SNe with higher velocities are typically more luminous than lower-velocity SNe, high-velocity SNe are more likely to be discovered at larger redshifts than low-velocity SNe. To explore this potential bias, we divide our sample into two groups, one at low redshift and the other at high redshift, and perform Kolmogorov-Smirnov (KS) test on the two groups. Figure~\ref{fig:z_histogram} (left panel) shows the histogram of all SNe~Ia in our sample that have host-galaxy redshift information from the kaepora database (291 out of 311 objects in our sample). We set the break redshift to be $z_b = \text{\numlist{0.01; 0.02; 0.03; 0.04}}$ (note that both the mean and median redshifts are 0.02). Tests give KS values of 0.13, 0.15, 0.14, and 0.13, and (respectively) $p$-values of 0.37, 0.05, 0.21, and 0.71; Figure~\ref{fig:z_histogram} (right panel) shows an example for $z_b = 0.03$. Our test demonstrates that except for $z_b = 0.02$, all other tests ($z_b = \text{\numlist{0.01; 0.03; 0.04}}$) did not exhibit a significant difference between the two groups. It is unknown why the KS test shows a significant difference between the two groups only at $z_b = 0.02$; so, in addition, we conducted a KS test with $z < 0.01$ (64 SNe) and $z > 0.04$ (30 SNe), giving a KS value of 0.16 and a $p$-value of 0.61, suggestive of insignificant differences. We thus conclude that there is no strong bias in our sample with SNe at low and high redshifts. 

\section{Bimodal Justification} \label{sec:bimodal_justifcation}

\cite{wang13} found that the distribution of \siII velocities near the time of peak brightness in SNe~Ia shows a bimodal structure (see their  Fig. 1$c$), with normal-velocity and high-velocity Gaussian components having respective peaks centred at \SIlist{10800; 13000}{\km\per\s} in the fit. While this bimodal structure is relatively obvious, they did not show the statistical improvement of the two-Gaussian fit over the one-Gaussian fit. Here we examine this issue with the above-selected sample of 311 SNe~Ia, nearly doubling the number of objects (165 were used by \citealt{wang13}). 

There are several methods for testing the significance in distributions, a popular one being the Pearson's $\chi^2$ statistic (not to be confused with reduced $\chi^2$ statistics). Pearson's $\chi^2$ test can handle cases with counts that are dimensionless and have no meaningful uncertainties, as compared with those used in reduced $\chi^2$ statistics where the errors come from the variance in each observation. A disadvantage of Pearson's $\chi^2$ method is that it depends on how one bins the data, and problems can be caused if the number of counts within some bins is small \citep[e.g.,][]{horvath98}. However, another popular technique, the maximum-likelihood (ML) method \citep[e.g.,][]{horvath08}, is not sensitive to this problem because it treats each measurement independently. For our purposes, given that the observed \siII velocities could be rare at the high- or low-velocity ends (and thus the counts could be small within those bins), it is more appropriate to adopt the ML method.

\subsection{Maximum-Likelihood Method} \label{sec:maximum_likelihood_method}

The ML method requires the assumption that each observation $x$ comes from an underlying population distribution modeled as a probability density function $g(x,p_1,\ldots,p_m)$, where $p_1,\ldots,p_m$ are parameters in the probability density function. Having $N$ observations of $x$, we assume each observation $x$ is independently sampled from identical distributions. Therefore, the chance of obtaining all $N$ observations is given by the function called the likelihood function,

\begin{equation}
l = \prod_{i=1}^{N}  g(x_i,p_1,\ldots,p_m) \,,
\end{equation}

\noindent or in a more convenient logarithmic form called the log-likelihood,

\begin{equation}
L = \log l = \sum_{i=1}^{N} \log  \left( g (x_i,p_1,\ldots,p_m) \right) \,.
\end{equation}

The ML procedure maximises the log-likelihood $L_\text{max}$ for all possible values of the parameter ${p_1,\ldots,p_m}$ determined by some defined parameter space. Since the logarithmic function is monotonic, the log-likelihood function reaches maximum where the likelihood function does as well.

Since the likelihood is the chance of an observation under some hypothesis of the underlying population, if there are two hypotheses, the ratio of two likelihoods (called the likelihood ratio) states how likely one hypothesis is  compared to the other. The likelihood ratio is better interpreted using the likelihood ratio $\chi^2$ statistics (not to be confused with the Pearson's $\chi^2$ statistics as introduced in the previous section),

\begin{equation}
\chi^2 = 2 ( L_\mathrm{1,max} -  L_\mathrm{0,max}) , \label{eq:chi_square_statistics}
\end{equation}

\noindent
where $L_\mathrm{0,max}$ and $L_\mathrm{1,max}$ are the log-likelihoods of the null hypothesis and alternative hypothesis, respectively. The likelihood ratio $\chi^2$ statistics is asymptotically distributed as $\chi^2$ with $d$ degrees of freedom, which equals the difference in the number of free parameters between the alternative and null hypothesis.

\subsection{Applying Maximum Likelihood to the Data} \label{sec:maximum_likelihood_method_apply_to_the_data}

In our case, the observable variable $x$ is the \siII velocity $v$. We take the hypothesis that $v$ is distributed normally; more specifically, it is distributed as the sum of $k$ normal components, where the components have the functional form

\begin{equation}
f_l(v,\mu_l, \sigma_l) = \frac{1}{\sigma_l \sqrt{2\pi}} \exp\left( -\frac{(v- \mu_l)^2}{2\sigma_l^2} \right) \,,
\label{eq:component}
\end{equation}

\noindent
where $\mu_l$ and $\sigma_l$ are  respectively the mean and standard deviation (unknown parameters). We therefore have the probability density  function

\begin{equation}
g(v_i,\mu_1, \sigma_1, \ldots, \mu_k, \sigma_k) = \sum_{l=1}^k w_l f(v_i,\mu_l,\sigma_l ) \,,
\label{eq:pdf}
\end{equation}

\noindent
where $w_l$ is the weight for the $l$-th component (out of $k$ components), and $\sum_{l=1}^k w_l = N$. The log-likelihood function is then composed of the logged sum across all probabilities for $N=311$ SNe,

\begin{equation}
L_k = \sum_{i=1}^N \log \left(\sum_{l=1}^k w_l f_l(v_i, \mu_l,\sigma_l ) \right). \label{eq:loglikelihood_gaussian}
\end{equation}

To find the maximum of $L_k$, we adopt SciPy's implementation of Nelder-Mead optimisation, \texttt{scipy.optimize}\footnote{\url{https://docs.scipy.org/doc/scipy/reference/optimize.html}}, in Python 3 to determine the argmax simultaneously for each of the parameters $\mu_l$, $\sigma_l$, and $w_l$. We start with only one normal component ($k = 1$) and find $L_\mathrm{1,max} = -2683$. We then add another normal component ($k = 2$) and find $L_\mathrm{2,max} = -2648$, statistically significantly higher than $L_\mathrm{1,max}$, which is a strong argument that two-component fitting is better than one-component fitting (see below for a more detailed discussion about the statistical improvement). We also try to adopt a third normal component, but the optimisation program was not able to converge and the resulting log-likelihood does not significantly differ from the results of using two components. This indicates that a third component is not needed for the fitting. We thus focus our discussion on the one- and two-component cases.

\begin{table*}
    \centering
    \caption{Best Parameters from ML and $\chi^2$ Fitting Methods}
    \label{tab:population_fitting}
    \begin{tabular}{cccccccc}
    \hline
    {} ML        & $L_\text{max}$ & $w_1$ &  $\mu_1$        &  $\sigma_1$      & $w_2$ & $\mu_2$        &  $\sigma_2$ \\
    {} method    &                &       &  \si{(\km\per\s)} & \si{(\km\per\s)}   &       & \si{(\km\per\s)} & \si{(\km\per\s)}   \\
    \hline
    $k=1$        & -2682       & 311   &  11,500        &  1300        &       &      {}      &  {}     \\
    $k=2$        & -2647       & 201   &  11,000        &  700         &  110  &  12,300       &  1800    \\
    \hline
    \\
    {} $\chi^2$  & $\chi^2$/dof   & $w_1$ &    $\mu_1$      &  $\sigma_1$      & $w_2$ & $\mu_2$        &  $\sigma_2$ \\
    {} method    &                &       &  \si{(\km\per\s)} & \si{(\km\per\s)}   &       & \si{(\km\per\s)} & \si{(\km\per\s)}   \\
    \hline
    $k=1$        &  18.7/8       & 311   &  11,100          &  1000             &       &      {}        &      {}     \\
    $k=2$        &  9.44/5         & 236   &  11,000          &  800            &  75   &  13,200        &  1400    \\
    \hline
    \end{tabular}
\end{table*}

\begin{figure*}
    \centering
    \includegraphics[width=\textwidth]{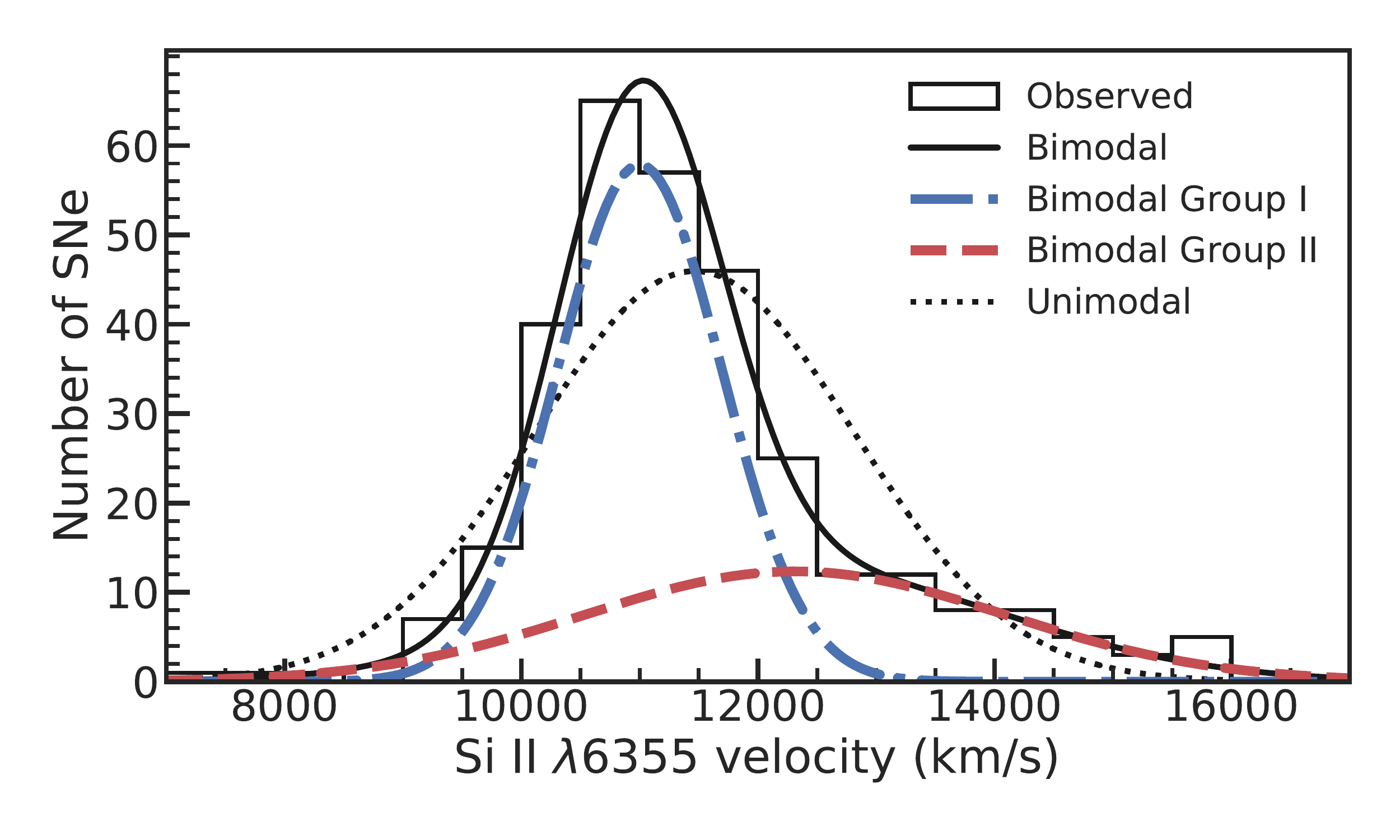}
    \caption{Histogram of the \siII velocity distribution of 311 SNe~Ia from the kaepora sample. Black solid line is the bimodal fit from the ML method, and black dotted line is the unimodal fit with parameters described in Table~\ref{tab:population_fitting}. The bimodal fit consists of two components: Group I SNe~Ia with lower and narrower-scatter velocity, $\mu_1 = \bimodalmuI$ and $\sigma_1 = \bimodalsigmaI$ (blue dot-dashed line), and Group II SNe~Ia with higher and broader-scatter velocity, $\mu_2 = \bimodalmuII$ and $\sigma_2 = \bimodalsigmaII$ (red dashed line).}
    \label{fig:fitted_velocity_dist}
\end{figure*}

In Table~\ref{tab:population_fitting}, we list the best estimated parameters for the one-  and two-component cases and plot the results in Figure~\ref{fig:fitted_velocity_dist}. In the one-component case, we find best-estimated parameters of $\mu_1 = \unimodalmu$ and $\sigma_1 = \unimodalsigma$ with $w_1$ fixed to be 311 (shown as the black dotted line in Figure~\ref{fig:fitted_velocity_dist}). The two-component case consists of a group of SNe~Ia with a lower and narrower-scatter velocity ($\mu_1 = \bimodalmuI$, $\sigma_1 = \bimodalsigmaI$, and $w_1=201$, shown as a blue dot-dashed line in Figure~\ref{fig:fitted_velocity_dist}), and another group of SNe~Ia with a higher and broader-scatter velocity ($\mu_2 = \bimodalmuII$, $\sigma_2 = \bimodalsigmaII$, and $w_2=110$, shown as a red dashed line in Figure~\ref{fig:fitted_velocity_dist}). The combination of the two groups is illustrated with a black solid line in Figure~\ref{fig:fitted_velocity_dist}. Our results are similar to those of \cite{wang13}, who reported a lower velocity with $\mu_1 = \SI{10800}{\km\per\s}$ and a higher velocity with $\mu_2 = \SI{13000}{\km\per\s}$ (no scatter values were reported).

For comparison, we also use the least-squares binned fitting method scored by the Pearson's $\chi^2$ statistic to fit the histogram distribution of \siII velocities with one and two normal components. The data are binned into 9 bins with width $\SI{500}{\km\per\s}$, where bin 1 is $v\le \SI{9500}{\km\per\s}$ and bin 9 is $v > \SI{13000}{\km\per\s}$. The bins were chosen purposefully so that each bin has at least an expected count of 4. The fitting results are also listed in Table \ref{tab:population_fitting}; we see that the best-fit parameters from the $\chi^2$ fitting are very close to those of the ML method. For the two-components case, the $\chi^2$ fitting also reveals that there is a lower velocity, narrower-scatter group and a higher velocity, broader scatter group, confirming the findings from the ML method.
Note that since Pearson's $\chi^2$ method has the disadvantage of depending on how one bins the data, we only adopt ML method result for further analysis.

\subsection{Statistical Improvement} \label{sec:statistical_improvements}

To estimate the statistical improvement of the two-component fitting over the one-component fitting, we first look at the likelihood ratio $\chi^2$ statistics in the ML method according to Equation~\eqref{eq:chi_square_statistics}. We take the null hypothesis that the data are distributed as a unimodal normal $k = 1$. The alternative hypothesis adds a new component --- that is, moving from $k = 1$ to $k = 2$, the ML solution of $L_\mathrm{1,max}$ changes to $L_\mathrm{2,max}$. This also increases the number of free parameters by 3 ($w_2$, $\mu_2$, and $\sigma_2)$. Applying Equation~\eqref{eq:chi_square_statistics} to $L_\mathrm{2,max}$ and $L_\mathrm{1,max}$, we find that the likelihood ratio $\chi^2$ statistic is 70 with 3 degrees of freedom. The $p$-value calculated from the $\chi^2$ statistic is much less than 0.01; hence, the improvement of the two-component model over the one-component model is statistically significant.

As an independent check, we also adopt the Akaike Information Criterion (AIC; \citealt{akaike74}) to apply a penalty according to the number of free parameters to prevent overfitting. The AIC makes it possible to perform model selection when the models have different numbers of free parameters. To be sensitive to the sampling error of a small dataset, we use the modified version of AIC denoted as AICc (\citealt{sugiura78}),

\begin{equation}
\text{AICc} = -2 L_\text{max} + 2d + \frac{2d(d + 1)}{N - d - 1} \,
\label{eq:aicc}
\end{equation}

\noindent
where $d$ is the number of parameters and $N$ is the sample size. A difference of 2 in the AICc provides positive evidence for the model having higher AICc, while a difference of 6 offers strong positive evidence (e.g., \citealt{Kass95, mukherjee98}). We use the $L_\text{max}$ from the likelihood fitting method given in Table 1. For $k = 1$ and $k  = 2$, we calculate $\text{AICc} = 5360$ and 5285, respectively; thus, from one component ($k=1$) to two components ($k=2$), the change in AICc is 75 (much greater than 6), which gives strong positive evidence that $k = 2$ produces a better fit than $k = 1$.

In conclusion, both the likelihood ratio test and the AICc test show strong statistical improvements for $k = 2$ fitting than $k = 1$ fitting. Therefore, if the velocity distribution is distributed normally (i.e., Gaussian mixture models), then the observed velocity is best described as two independent Gaussian distributions (Group I and Group II): one group with lower, narrower-scatter velocity ($\mu_1 = \bimodalmuI$, $\sigma_1 = \bimodalsigmaI$) and another group with higher, broader-scatter velocity ($\mu_2 = \bimodalmuII$, $\sigma_2 = \bimodalsigmaII$).
The ratio of the two groups is 201:110 (corresponding to 65\%:35\%) in the number of samples from the weight parameters. This velocity distribution is henceforth called the bimodal Gaussian distribution.

As one can see from Figure~\ref{fig:fitted_velocity_dist} for the two-component result, the second group of SNe~Ia, whose mean velocity is higher than that of the first group, also has greater scatter than the first group, so there is a large overlap with the first group on the low-velocity side (the high-velocity side is dominated by the second group). In fact, the low-velocity side from the second group extends even farther than the first group. We thus think it is inappropriate to name the first group as low-velocity SNe~Ia and the second group as high-velocity SNe~Ia. Instead, here we suggest naming the lower mean velocity group as ``Group I" and the other group as ``Group II."

Because of the large overlap between the two groups, it is difficult to assign a specific SN~Ia with a given velocity to a specific group, especially if the velocity is in the range \SIrange{9500}{12000}{\km\per\s}. But for a velocity beyond that range, one can compare the probability between the two groups according to the probability distribution function assuming the \siII velocity data came from independently sampling each (but be cautious because the observed SN
counts in those ranges are small). For $v < \SI{9500}{\km\per\s}$, the probability ratio between Group I and Group II is 21\%:79\%; for $v > \SI{12000}{\km\per\s}$, the  ratio is 12\%:88\%. Furthermore, for $v >$ \SI{13000}{\km\per\s}, the probability for being in Group II is $>99\%$, which means almost all SNe~Ia with velocity $>$ \SI{13000}{\km\per\s} belong to Group II according to the fitting.

\section{Statistical Simulation of the Velocity Distribution}

As we demonstrated above, it is almost certain that  SN~Ia ejecta velocities exhibit a bimodal Gaussian distribution: Group I with lower and narrower-scatter velocities and Group II with higher and broader-scatter velocities. Some recent studies show that these two groups have different properties. For example,
polarization observations reveal that the high-velocity SNe~Ia (belonging to Group II) tend to have larger line polarizations \citep{maund10,cikota19}.
\cite{wang13} found that high-velocity SNe~Ia are substantially more concentrated in the inner and brighter regions of their host galaxies.
\cite{wang19} also found that high-velocity SNe~Ia tend to have cold, dusty circumstellar material around their progenitors, based on the variable Na~I~D absorption and flattening of the $B$-band light curve starting 40\,d after maximum light.
\cite{zheng18} found that high-velocity SNe~Ia are generally not good standardisable candles. Furthermore, \cite{polin19} proposed a model for a subclass of SNe~Ia from sub-Chandrasekhar-mass progenitors with thin helium shells, and use the data from \cite{zheng18} to show that this model could be consistent with Group II SNe~Ia.

While the true origin of the two components remains unknown, one simple explanation could be that there exist two intrinsic velocity distributions as we observed. However, here we focus on exploring another possibility, asymmetric geometry. We assume that all SNe~Ia share the same intrinsic velocity distributions and test whether the observed bimodal velocity distributions could be caused by asymmetry for some SNe~Ia. Our goal is to use asymmetric geometric models through statistical simulations to see whether the velocity distribution of Group II SNe~Ia could be caused by angular geometry effects (i.e., ejecta-velocity variations at different viewing angles). Such asymmetries could be produced by various explosion models, such as delayed-detonation  \citep[e.g.,][]{seitenzahl13,sim13}, detonation from failed deflagration \citep[e.g.,][]{kasen07}, double detonation \citep[e.g.,][]{townsley19}, violent white dwarf mergers \citep[e.g.,][]{pakmor10, moll14,raskin14}, and others \citep[e.g.,][]{kasen09}. The ejecta velocities measured from different viewing angles could differ by up to \SI{4000}{\km\per\s} \citep[e.g.,][]{kasen07, kasen09, townsley19, levanon19}.

\subsection{Two Asymmetric Models}

Before starting the simulation of the velocity distributions, we need to quantify the velocity changes as a function of the viewing angle due to the asymmetric ejecta. To do this, we use the model spectra synthesised at different viewing angles to measure the \siII velocity. We adopt the synthesised spectra from two different models, one from \cite{kasen07} and the other from \cite{townsley19}; these models provide  spectra at the time of SN~Ia peak brightness for different viewing angles. Note, however, that our simulation setups can be updated  easily to adopt other models, if desired.

\begin{figure*}
    \centering
    \includegraphics[width=\textwidth]{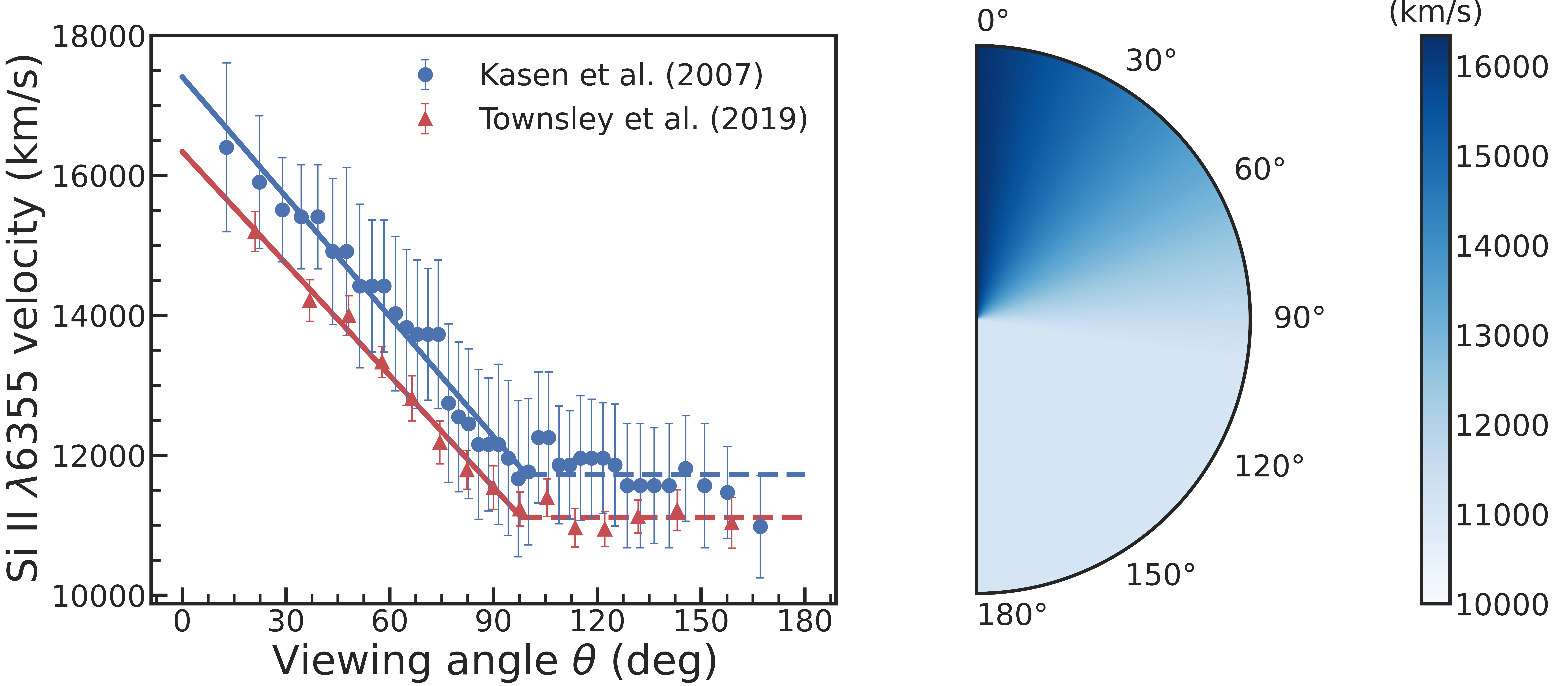}
    \caption{Left: \siII velocity (measured from the maximum-light synthetic spectra) as a function of viewing angle from the \citet{kasen07} and \citet{townsley19} models. The data are fitted with a monotonically decreasing function as described in the text. Right: Demonstration of the ejecta seen at different viewing angles which are color coded for the different velocities.}
    \label{fig:kasen07_fitted}
\end{figure*}

\citet{kasen07} studied in detail one model called Y12, selected from the detonating failed deflagration (DFD) scenario \citep{plewa07}, which considers an off-center, mild ignition process in a degenerate Chandrasekhar-mass C-O white dwarf. In the Y12 model, the white dwarf is initially ignited within a small spherical region on the axis of symmetry, 50\,km in size and offset 12.5\,km from the center. A temporal series  of synthetic spectra was calculated for different viewing angles; of particular interest to us are the  spectra near the time of maximum brightness  (see Fig. 7 of \citealt{kasen07}). In this model, the viewing angle was defined as $\theta = \ang{0}$ on the ignition side and $\theta = \ang{180}$ on the detonation side. We measure the \siII velocity from the maximum-light synthetic spectra at different viewing angles (simply treating them as real spectra) and plot the results in Figure~\ref{fig:kasen07_fitted}, shown as blue circles.

\cite{townsley19} considered a double detonation that ignites first in a helium shell on the surface of the white dwarf; the He shell is heated by a directly impacting accretion stream and mixes modestly with the outer edge of the core. They use a white dwarf with a \SI{1.0}{M_\odot} C-O core and a surface He layer with base $\rho=\SI{2e5}{\g\per\cm\cubed}$ and $T=\SI{5e8}{\K}$, having a mass of \SI{0.021}{M_\odot}. The He detonation is ignited with a spherical hotspot placed on the symmetry axis at the base of the He layer. Maximum-light spectra are synthesised at different averaged viewing angles (see \ref{fig:kasen07_fitted} of \citealt{townsley19}). Again, we measure the \siII velocity and plot the results in Figure~\ref{fig:kasen07_fitted}, shown as red circles.

One can see that both models give very similar behavior for the \siII velocity as a function of viewing angle: the highest velocity appears at $\theta = \ang{0}$. The velocity decreases linearly as the viewing angle increases (toward the detonation side). But after $\theta$ around \ang{100}, near the equator, the velocity becomes constant with almost no changes. To mathematically quantify this evolution, we use the following monotonically decreasing function to describe the data:

\begin{equation}
    v =
        \begin{cases}
            v_c + \frac{\theta_c - \theta}{\theta_c}\Delta v    &     \theta \leq \theta_c    \\
            v_c                                                 &     \theta > \theta_c,
        \end{cases}
        \label{eq:angle_velocity_model}
\end{equation}

\noindent
where $\theta_c$ is the cutoff viewing angle, $v_c$ is the constant velocity after the cutoff viewing angle, and $\Delta v$ is the velocity difference between the highest velocity (where $\theta = \ang{0}$) and $v_c$. We use this function to fit the data from both models and plot the results in Figure~\ref{fig:kasen07_fitted}. We find $\theta_c = \SI{100 \pm 8}{\degree}$, $v_c = \SI{11,700 \pm 200}{\km\per\s}$, and $\Delta v = \SI{5700 \pm 990}{\km\per\s}$ for the \cite{kasen07} model, and $\theta_c = \SI{98 \pm 4}{\degree}$, $v_c = \SI{11,100 \pm 110}{\km\per\s}$, and $\Delta v = \SI{5200 \pm 430}{\km\per\s}$ for the \cite{townsley19} model. The values for these parameters are very close to each other among the two models. Because of this, we will adopt only $\theta_c$, $v_c$, and $\Delta v$ as a single set of free parameters to be determined by the statistical simulation best fitted to the data as described in Section \ref{sec:simulation}. 

\subsection{Simulation Setups} \label{sec:simulation}

To simulate the distribution of velocities, we generate a sample size of $N = \num{10000}$ SN~Ia cutoff velocities $v_c$ individually sampled from the Group I Gaussian distribution, henceforth calling this sample the intrinsic distribution, which is also the input distribution for our simulation. We then consider the asymmetry effect according to the different parameters, applying it to a certain fraction of the SNe~Ia (see details below); thus, the asymmetry-corrected velocity distribution, which is also the output sample from our simulation, changes from the intrinsic (input) distribution. We finally compare the asymmetry-corrected (output) velocity distribution with the observed distribution by adopting the KS test in order to find the best parameters. We define the parameter space from a grid search allowing the parameters to be within their respective ranges: $\theta_c \in \set{\ang{0}, \ang{1}, \ang{2}, \ldots, \ang{180}}$ with an increment of \ang{1}, and $\Delta v \in \set{3000, 3100, 3200, \ldots, 7500}\si{\km\per\s}$ with an increment of \SI{100}{\km\per\s}, giving a space of size 8280 pairs of parameters. The parameters are ranked by the KS values, and we select the top four best parameters for further discussion from the total of 8280 pairs of parameters. The simulation codes are written in Python 3 and can be found online. \footnote{\url{https://github.com/ketozhang/ejecta_velocities_of_type_Ia_supernovae}}.

\subsection{Results} \label{sec:simresult}

We consider two cases for the asymmetry effect in our simulation: (1) only a portion of the SNe~Ia from the intrinsic distribution are asymmetric, and (2) all of the SNe~Ia from the intrinsic distribution are asymmetric.

\begin{figure*}
    \centering
    \includegraphics[width=\textwidth]{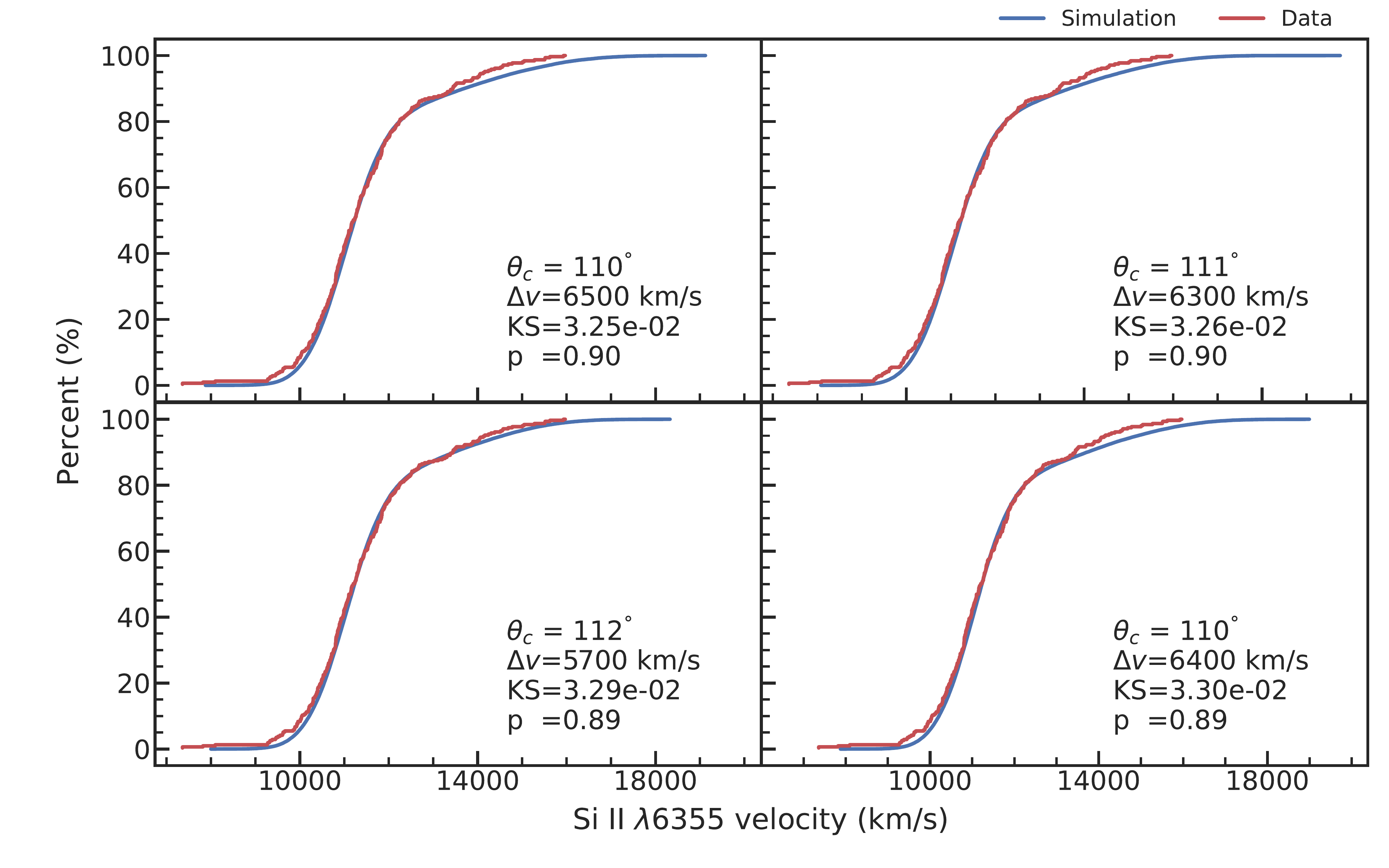}
    \caption{Cumulative distributions of the top four best-performing simulations (selected by the KS statistic) for Case 1, where a portion (35\%) of all SNe~Ia are assumed to be asymmetric in the simulation. These four simulations (blue) can reproduce the observed (red) velocity distribution.}
    \label{fig:grid_search_cumulative_results_v2}
\end{figure*}

For Case 1, considering the mean and scatter parameters of the two groups, it is natural to assume that Group I SNe~Ia are symmetric while Group II SNe~Ia are asymmetric, and we therefore assume 35\% of the intrinsic distribution (namely the total number of Group II SNe~Ia) need to be corrected for asymmetry from the intrinsic (input) distribution. To do this, we first generate $N = 10^5$ SN~Ia velocities sampled from a Gaussian distribution with parameters taken from Group I SNe~Ia ($\mu = \bimodalmuI$ and $\sigma = \bimodalsigmaI$). 64.6\% of these SNe~Ia (namely the total number in Group I) are unmodified and passed onto the final output result. For the remaining 35\% of SNe~Ia, being Group II, we treat them by applying the asymmetry model (Eq. \ref{eq:angle_velocity_model}), where $v_c$ is taken from the generated velocities of the intrinsic distribution (i.e., $v_c$ is sampled from the Group I Gaussian distribution), while $\theta_c$ and $\Delta_v$ are taken from the parameter space. The asymmetry-corrected (output) velocity for the final result is then calculated assuming the SN is equally observable in each unit solid angle. We achieve this by applying the spherical point-picking correction to the viewing angle $\theta = \arccos(2u-1)$, where $u$ samples the uniform$(0\,, 1)$ distribution. We finally compare the asymmetry-corrected velocity distribution for all $10^5$ SNe~Ia with the observed distribution of 311 SNe~Ia by adopting the KS test. This step is repeated for a total of 8280 sets of parameters $\theta_c$ and $\Delta v$ in the parameter space.

Here we show the top four performing parameter sets in the KS test, as plotted in Figure~\ref{fig:grid_search_cumulative_results_v2}. The KS statistic computes a distance between the empirical distribution of the simulation of $10^5$ SNe~Ia and the observed sample of 311 SNe~Ia. The $p$-value is two-tailed, generated from the KS statistic and the null hypothesis that the two distributions come from the same population distribution. For reference, the KS statistic with the naive hypothesis that the underlying distribution is Gaussian with Group I parameters tested with the observed data produces a KS value of $0.19$ and a $p$-value of $\num{6.43e-10}$; hence, we strongly reject the hypothesis that the observed data are normally distributed with Group I parameters. For the top four performing results shown in Figure~\ref{fig:grid_search_cumulative_results_v2}, with an $\alpha$ level of 5\%, the null hypothesis cannot be rejected. The best parameters out of the top four cases are $\theta_c = \ang{111}$ and $\Delta v = \SI{6000}{\km\per\s}$ with a KS value of $\num{3.2e-02}$ and a $p$-value of $0.91$. All four top cases have very similar best parameters, with $\theta_c$ in the range \SIrange{110}{112}{\degree} and $\Delta v$ of \SIrange{5900}{6300}{\km\per\s}. These parameters are very close to the values from the two asymmetric models given in Section 4.1. To produce confidence intervals, we create a joint probability space where the values are the parameters and the chance to observe the parameters is the $p$-value normalised by dividing the $p$-value with the sum of all 8280 $p$-values from the simulations. This probability space is then sampled \num{10000} times to get a parameter distribution where the mean and standard deviation are used for the $1\sigma$ confidence interval. Doing so, the $1\sigma$ confidence interval is $\theta_c = \caseItheta$ and $\Delta v = \caseIdeltav$ at 1$\sigma$ confidence.

\begin{figure*}
    \centering
    \includegraphics[width=\textwidth]{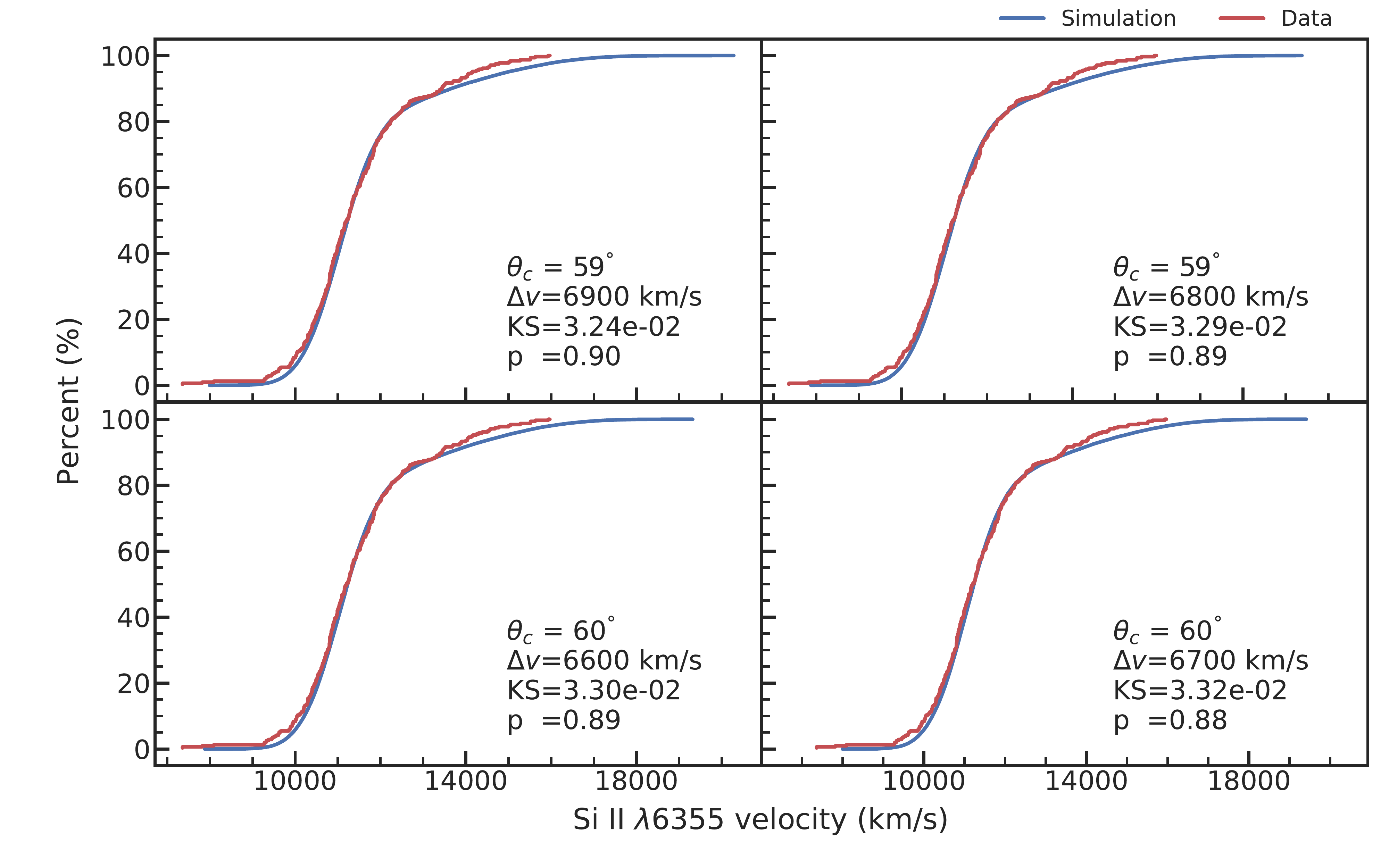}
    \caption{Same as Figure~\ref{fig:grid_search_cumulative_results_v2} but for Case 2, where all of the SNe~Ia are assumed to be asymmetric in the simulation. These top four cases also can reproduce the observed (red) velocity distribution. Compared to Case 1, the obvious difference is the $\theta_c$ parameter, which decreased substantially.}
    \label{fig:grid_search_cumulative_results_v3}
\end{figure*}

For Case 2, instead of applying the asymmetric model to a subset of the intrinsic (input) distribution, we apply the asymmetric model to all $N = \num{10000}$ simulated SNe~Ia from the intrinsic (input) distribution. Thus, we assume that all SNe~Ia are asymmetric. The remaining simulation procedures follow exactly as in the Case 1. Similarly, we show the top four performing results in Figure~\ref{fig:grid_search_cumulative_results_v3}. At an $\alpha$ level of 5\%, the null hypothesis for all four top cases cannot be rejected, the same as in Case 1. The best parameters out of the top four simulations are $\theta_c = \ang{60}$ and $\Delta v = \SI{6400}{\km\per\s}$, with a KS value of $\num{3.3e-02}$ and a $p$-value of $0.89$. The $1\sigma$ confidence interval is calculated to be $\theta_c = \caseIItheta$ and $\Delta v = \caseIIdeltav$. Compared to Case 1, the obvious difference is in $\theta_c$: for Case 2,  $\theta_c$ is around $\ang{63}$, which is much smaller than in Case 1 ($\theta_c \approx \ang{120}$). The other parameter, $\Delta v$, is almost the same as for Case 1.

For both cases, at an $\alpha$ level of 5\%, the null hypothesis for all four top simulations cannot be rejected, which indicates that for both cases, the simulations can reproduce the observed \siII velocity distribution. Also, while comparing to the models, both cases give a cutoff velocity $\Delta v$ around $\SI{5500}{\km\per\s}$, matching well to the two models ($\SI{5200}{\km\per\s}$ and $\SI{5700}{\km\per\s}$). However, for the other parameter, the cutoff angle $\theta_c$, Case 1 is apparently better than Case 2. The $\theta_c$ parameter from both models is $\ang{\sim 100}$. From Case 1, $\theta_c$ is $\ang{\sim 120}$, slightly larger, but much closer to $\ang{100}$ compared to Case 2, for which $\theta_c$ is $\ang{\sim 63}$. For this reason, we favor Case 1 over Case 2.

As shown above, our simulations indicate that assuming a portion of SNe~Ia are asymmetric, we can reproduce the observed \siII velocity distribution (Case 1). The asymmetric SNe~Ia could occupy a notable fraction (35\%) of the total SNe~Ia. In fact, since the $\theta_c$ value in Case 1 $\ang{120 \pm 15}$ is slightly larger than in the model ($\ang{\sim 100}$), while increasing the fraction of asymmetric SNe~Ia in the simulation would reduce the $\theta_c$ value, a larger fraction of asymmetric SNe~Ia would provide a better match between the simulation $\theta_c$ and the model; therefore our simulation indicates that more than 35\% of SNe~Ia are asymmetric.
Note that for both cases, since the input distribution is the same for all SNe, all SNe~Ia share the same intrinsic velocity distribution in our simulation. Even for Case 1, where 35\% are asymmetric,
all SNe~Ia still share the same input distribution. The only difference is that the output distribution for the 35\% SNe Ia are asymmetry-corrected.
We therefore conclude that more than 35\% of SNe~Ia are asymmetric from our simulation if all SNe~Ia share the same intrinsic velocity distribution based on the adopted models and simulations.

\section{Conclusions and Discussion}

We have studied the ejecta velocity distribution of a sample of 311 SNe~Ia derived from the \siII absorption line measured at the time of peak brightness. We statistically show that the bimodal Gaussian yields the best fit in maximum likelihood to the observed velocity distribution, significantly better than one-Gaussian fitting. The bimodality suggests two velocity groups among SNe~Ia:
Group I with a lower velocity but narrow scatter ($\mu_1 = \bimodalmuI$, $\sigma_1 = \bimodalsigmaI$), and Group II with a higher velocity but broader  scatter ($\mu_2 = \bimodalmuII$, $\sigma_2 = \bimodalsigmaII$). The number ratio of the two groups is 201:110 (65\%:35\%) in our sample. There is substantial degeneracy between the two groups for SNe~Ia with ejecta velocity below $\SI{12000}{\km\per\s}$. However, for SNe~Ia with velocity exceeding this value, the distribution is dominated by Group II (with a probability of 12\%:88\% between Group I and Group II), and almost all SNe~Ia with velocity $> \SI{13000}{\km\per\s}$ belong to Group II (the probability of being in Group II is $>99\%$).

Although the true origin of the two components is unknown, one simple explanation could be that naturally there exist two intrinsic velocity distributions as observed. However, we try to use asymmetric geometric models through statistical simulations to reproduce the observed velocity distributions assuming all SNe~Ia originate from the same intrinsic (input) distribution.
Specifically, we adopt a geometric model in which the velocity decreases linearly as the viewing angle $\theta$ increases, but then becomes constant after a certain cutoff $\theta_c$. The velocity difference between $\theta = \ang{0}$ and $\theta_c$ gives another $\Delta v$ parameter in our simulation. A total of 8280 pairs of $\theta$ and $\Delta v$ are set up in our simulation. For each pair, we generate a sample size of $N = 10^5$ velocities from the assumed intrinsic Gaussian distribution, and then we apply the geometric models to a certain fraction of them. We finally compare the asymmetry-corrected velocity (output) distribution with the observed distribution by adopting the KS test to find the best values of $\theta$ and $\Delta v$. 

We considered two cases in our simulation. In Case 1, where only a portion (35\%) of all SNe~Ia are asymmetric, our simulation yields best-fit parameters $\theta_c = \caseItheta$ and $\Delta v = \caseIdeltav$. In Case 2, where all SNe~Ia are asymmetric, our simulation yields best-fit parameters $\theta_c = \caseIItheta$ and $\Delta v = \caseIIdeltav$. We find that for both cases, the simulations can reproduce the observed \siII velocity distribution, giving a cutoff velocity $\Delta v \approx \SI{5500}{\km\per\s}$, matching the models well. However, considering the other parameter (the cutoff angle $\theta_c$), Case 1 $\caseItheta$ is better than Case 2 $\caseIItheta$ compared with the models ($\ang{\sim100}$), so we favor Case 1. Regardless, we find that a significantly large portion ($>$35\%) of SNe~Ia are asymmetric in our simulations, assuming all SNe~Ia originate from the same explosion mechanism.

Observationally, spectropolarimetry provides an effective way to explore the geometry of SNe~Ia (see \citealt{wang08} for a review).
Interestingly, recent observations show that the \siII line polarization is correlated with the velocity of that line \citep{maund10, cikota19}: SNe~Ia with higher \siII velocity tend to have stronger polarization. This correlation is consistent with our simulation Case 1, that a portion (35\%) of all SNe~Ia are asymmetric (and thus should have higher polarization), resulting in Group II SNe~Ia with higher \siII velocity on average. This provides another indication that simulation Case 1 is more favoured over Case 2.
However, although our simulations show that a large portion of SNe~Ia are asymmetric, many models can result in asymmetric explosions, so we are unable to distinguish which model is more favoured. Additional observations are required to explore more specific explosion models. For example, using high-quality nebular spectra of SNe~Ia, \cite{dong15} discovered clear double-peaked line profiles in 3 out of $\sim 20$ SNe~Ia,
which is naturally expected from the direct white dwarf collision model \citep[e.g.,][]{kushnir13}, or the violent merger model when the companion is completely disrupted \citep[e.g.,][]{pakmor12}.

We therefore encourage more polarization and nebular observations of SNe~Ia for further studies of asymmetries.

\section*{Acknowledgements}

We thank the anonymous referee for useful comments. K.D.Z. was supported by the Anslem M\&PS Fund through the Berkeley Summer Undergraduate Research Fellowship (SURF) program in 2019. A.V.F.'s group at U.C. Berkeley is grateful for financial assistance from Gary \& Cynthia Bengier (T.de J. is a Bengier Postdoctoral Fellow), Marc J. Staley (B.E.S. is a Marc J. Staley Graduate Fellow), the Christopher R. Redlich Fund, the TABASGO Foundation, and the Miller Institute for Basic Research in Science (U.C. Berkeley). Research at Lick Observatory (where many of the SN~Ia spectra used here were obtained) is partially supported by a generous gift from Google, and K.D.Z. is a Google Lick Predoctoral Fellow.

\section{Data Availability}
The data and results underlying this article are available in Zenodo at \href{https://zenodo.org/record/4032100}{https://zenodo.org/record/4032100}.




\bibliographystyle{mnras}
\bibliography{references}



\appendix
\label{sec:appendix}




\bsp	
\label{lastpage}
\end{document}